\documentstyle[12pt]{article}

\begin{document}
\newcommand{\dia}{\begin{displaymath}}
\newcommand{\die}{\end{displaymath}}
\newcommand{\eqa}{\begin{equation}}
\newcommand{\eqe}{\end{equation}}
\newcommand{\eqna}{\begin{eqnarray}}
\newcommand{\eqne}{\end{eqnarray}}
\newcommand{\eqnaa}{\begin{eqnarray*}}
\newcommand{\eqnae}{\end{eqnarray*}}
\newcommand{\fraz}{\frac{1}{2}}
\newcommand{\frav}{\frac{1}{4}}
\newcommand{\frasz}{\frac{1}{\sqrt{2}}}
\newcommand{\tr}[1]{\mbox{Tr}\left\{#1\right\}}
\renewcommand{\theequation}{\thesection .\arabic{equation}}
\newcommand{\sect}{\setcounter{equation}{0}\section}
\newcommand{\labe}[1]{\label{#1}}

\title{An angle representation of QCD
\thanks{Invited talk presented at the workshop on
"Quantum Infrared Physics", Paris June 6--10, 1994}
}
\author{Dieter Stoll\\ Department of Physics,
University of Tokyo\\ Bunkyo--ku, Tokyo 113, Japan}
\date{  }
\maketitle
\begin{abstract}
For the sake of eliminating gauge variant degrees of freedom we discuss
the way to introduce angular variables in the hamiltonian formulation
of QCD. On the basis of an analysis of Gauss' law constraints
a particular choice is made for the variable transformation from gauge fields
to angular field variables. The resulting formulation is analogous to the one
of Bars in terms of corner variables and it is closely related to the
hamiltonian lattice QCD formulation.
Therefore the corner or angle formulation
may constitute an useful starting point for the investigation of the low
energy properties of QCD in terms of gauge invariant degrees of freedom.
\end{abstract}


\sect{Introduction}
One of the long standing problems in contemporary physics is understanding
confinement of quarks and gluons from first principles.
The difficulty in dealing with
the infrared properties of QCD is on the one hand due to the non--linear
gluonic interaction and on the other due to the constraints on the dynamics
of the fundamental degrees of freedom which originate from the requirement of
gauge invariance. In spite of the general belief that the non--linear
interaction gives rise to confinement it has been conjectured recently that
in fact the non--abelian constraints may be most important
\cite{Johnson}. Aiming at an understanding of the low energy properties
of QCD we should therefore try to develop approximations to the full QCD
dynamics after the gauge variant degrees of freedom have been identified and
isolated.

The aforementioned constraints are given by Gauss' law operators, which
generate a compact group in each point in space, telling us
that the gauge variant degrees of freedom are "angle" variables.
Due to the fact that field theory deals with an infinite number of degrees of
freedom, the choice of unphysical variables is to
a large extent arbitrary. Therefore various decompositions into unphysical
"angle" variables and remaining physical variables are possible to arrive
at the desired separation of unphysical degrees of freedom
\cite{Johnson,Goldstone,Lenz}. Although successful in that respect the
variables
chosen in this way to parametrize the physical Hilbert space may be inadequate
to account in a simple way for the dynamics relevant for the low energy
properties of QCD. For further variable changes, on the other hand, the
complexity of the so derived hamiltonians constitutes a basic obstacle.

Therefore our starting point is the assumption that not only the unphysical
but all variables are "angle" variables and the hamiltonian should be
expressed first in terms of these angular degrees of freedom before making
a separation into gauge variant and gauge invariant ones. To find a definition
of "angle" variables
in terms of gauge or electric fields we concentrate on an
analysis of Gauss' law operators. It will be shown that the form of these
operators suggests the introduction of "angles" which are non--locally
related to the gauge fields. By a variable transformation the originally
quantized gauge fields and  electric fields in the Hamiltonian can be
replaced by "angle" and angular momentum operators
respectively. The resulting formulation is analogous to the one in terms of
corner variables obtained by Bars \cite{Bars}. In contrast to similar
approaches \cite{Johnson,Goldstone} the separation into
gauge variant and gauge invariant degrees of freedom is not made
from the outset in the "angle" or corner variable formulation. Therefore
it may constitute an useful starting point for the search of approximations
to the full QCD dynamics intended to understand its nonperturbative aspects.

\sect{The hamiltonian in terms of angle variables}
We consider a Hamiltonian formulation of SU(N) gauge theories on
a d--dimensional torus. Choosing the Weyl gauge
$A_0=0$ we have the following Hamiltonian density
\dia {\cal H}=\sum_i\bar\psi(x)\gamma_i\left(i\partial_i+gA_i\right)\psi(x)
+m\bar\psi(x)\psi(x) +\fraz \sum_i E_i^a(x)E_i^a(x)+
\fraz \sum_{ij} \tr{F_{ij}F^{ij}}\ .\die
Imposing periodic boundary conditions for the gauge and anti--periodic ones for
the fermion fields we quantize canonically  ($E_i^a(x)= \partial_0A_i^a(x)$)
\eqnaa \left[E_i^a(x),A_j^b(y)\right]&=&-i\delta_{a,b} \delta_{ij}
\delta^d(x-y);\quad
a,b=1,\dots,N^2-1;\ i,j=1,\dots,d\\ \left\{\psi_{k,\alpha}^{\dag}(x),\psi_{l,
\beta}(y)\right\}&=&\delta_{k,l}\delta_{\alpha,\beta}\delta^d(x-y);\quad
k,l=1,\dots,N;\ \alpha,\beta=\mbox{Spinor indices}\eqnae
where it is understood that the $\delta$--functions are periodic, as well.
Since we have not fixed the gauge classically, Gauss' law operator is the
quantum mechanical generator of the gauge symmetry. It commutes with the
Hamiltonian and therefore physical, i.e. gauge invariant, eigenstates must
be annihilated by the generators of the symmetry
\eqna \left.G^a(x)\right|phys.> &=& 0\ \labe{constr}\\
G^a(x) &=& \sum_i\left[ \partial_iE_i^a(x)+gf^{abc}A_i^b(x)E_i^c(x) \right]
+g\psi^{\dag}(x)\frac{\lambda^a}{2}\psi(x) \label{fabc} \\
\left[G^a(x),G^b(y)\right] &=& igf^{abc}G^c(x)\delta^d(x-y) \labe{algebra}
\ . \eqne
Since the generators obey the Lie algebra of the gauge group
it is understood that out of $d\cdot (N^2-1)$ gauge degrees of freedom
only a set of $N^2-1$ "angle" variables in each point in space is changed
by gauge transformations . Consequently the constraints are satisfied if these
"angles" have been identified and Gauss' law operator has been transformed
such that it is the angular momentum operator only with respect to these
unphysical "angles". Physical states then correspond to s--wave states which
are annihilated by these angular momentum operators and the Hamiltonian after
transformation  will not contain the unphysical variables anymore.

Since we want to replace gauge and electric fields by "angles" and angular
momenta in such a way that the constraints can easily be implemented, we study
the form of Gauss' law operators first in 1+1 dimensions. In this case
the contribution in eq.(\ref{fabc})
\eqa f^{abc}A^b(x)E^c(x)\labe{cross}\eqe
acts locally as an angular momentum operator on $N(N-1)$ "angle" variables in
either the gauge field or the elctric field representation. The missing
$(N-1)$ "angle" variables could not be identified, if this was the complete
Gauss' law operator already. Therefore we must conclude that in 1+1 dimensions
the full number of $(N^2-1)$ variables in each point in space can only be
eliminated due to the presence of $\partial_xE(x)$ in the Gauss' law
operators. This term not only distinguishes the gauge
fields as source of the additional unphysical variables but also
introduces a non-locality into the Gauss' law operators.
Therefore it seems natural to assume that the "angle" variables
which are unphysical are nonlocally related to the gauge field variables.
Although this argument is rigorous only in 1+1 dimensions
we assume it to be an useful hypothesis for introducing "angle" variables
in any dimensions.

An expression for the gauge fields satisfying this
requirement is\footnote{Note that throughout the paper spatial indices are
not summed over unless explicitly indicated.}
\eqna A_i(x) &=& \frac{i}{g}V_i(x) \partial_i V_i^{\dag}(x) \quad (\mbox{no
summation})\labe{afeld}\\
V_i(x) &=& P\exp\left[ig\int_0^{x_i}dz_iA_i(x_i^\perp,z_i) \right]\labe{vdef}\\
V_i(x) &=& \exp\left[i\xi_i(x)\right],\quad \xi_i(x)=\xi_i^a(x)\frac{
\lambda^a}{2};\quad 0<x_i\leq L \ \labe{xidef}\eqne
where $V_i(x)$ is a SU(N) matrix parametrized in
terms of "angles" $\xi_i^a(x)$, P denotes path ordering and $x_i^\perp$
stands for all coordinates orthogonal to $x_i$. Since this definition together
with the specific choice of paths in eq.(\ref{vdef}) leads to an unique
relation\footnote{With the choice \protect $0<x_i\leq L$ the
"angles" are uniquely determined from gauge fields by eq.(\ref{vdef}) if the
derivative is taken from one side only \protect{
$\partial_if(x_i) = \lim_{\epsilon\rightarrow 0} \frac{1}{\epsilon}\left[
f(x_i)-f(x_i-\epsilon)\right] ;\quad 0<x_i\leq L \labe{deriv}$}.
In this way it is possible to work with periodic, although not continuous
"angles" and angular momenta.} between $\xi_i(x)$ and $A_i(x)$ a
change of variables from $A_i(x), E_i(x)$ to $\xi_i(x)$ and the corresponding
angular momenta $J_i(x)$ becomes feasible. Using eq.(\ref{afeld})
we rewrite fermionic and magnetic part of the hamiltonian
\eqna  \bar\psi(x)\left\{\gamma_i\left[i\partial_i+gA_i(x)\right] +m\right\}
\psi(x)= \left[\bar\psi(x)U_i(x)\right] \left\{ \gamma_ii\partial_i +m\right\}
\left[U_i^{\dag}(x) \psi(x)\right] \labe{ferm}\ ,\\
\left[D_i,D_j\right] = \left[i\partial_i+gA_i,i\partial_j+gA_j\right] =
-U_i\left\{ \partial_i\left[\left(U_i^{\dag}U_j\right)\partial_j
\left(U_j^{\dag}U_i \right)\right] \right\}U_i^{\dag}\ , \\ \Rightarrow
\tr{F_{ij}F_{ij}} = \frac{-1}{g^2}\tr{\partial_i\left[\left[
\left(U_i^{\dag}U_j\right) \partial_j\left(U_j^{\dag}U_i\right)\right]\right]
\left[\partial_i\left[\left( U_i^{\dag}  U_j\right)\partial_j
\left(U_j^{\dag}U_i\right)\right]\right]^{\dag}}\nonumber  . \eqne
In order to reformulate the electric part of the Hamiltonian which contains the
conjugate momenta of the gauge fields, we introduce the angular momentum
operators $J_k^c(z)$ \cite{Stoll1}. These operators
generate translations in the space of "angles" $\xi_k$ as may be
seen from the commutation relations
\eqna \left[J_i^a(x),V_j(y)\right]&=&\delta_{i,j}\delta^d(x-y)V_j(y)
\frac{\lambda^a}{2}\labe{jvcom}\\   \left[J_i^a(x),J_j^b(y)
\right]&=&\delta_{i,j}if^{abc}J_i^c(x)\delta^d(x-y)\ .\nonumber \eqne
We note that due to the periodicity of $V_i,J_i$ the $\delta$--functions
in these expressions are periodic, as well. Introducing furthermore the
orthogonal matrices $N_i$
\eqna N_i^{ac}(x) &=& \tr{V_i^{\dag}(x)\frac{\lambda^a}{2}V_i(x)\lambda^c}
\labe{ndef} \\  \left[J_i^b(z),N_j^{ac}(x)
\right]&=& if^{bce}N_j^{ae}(x)\delta_{i,j}\delta^d(x-z) \nonumber \eqne
one can derive the following expression
for the electric part of the hamiltonian density \cite{Stoll1}
\eqna E_i^a(x) &=& g\int d^dz \delta^{d-1}(z_i^\perp-x_i^\perp)\theta(z_i-x_i)
\theta(x_i)N_i^{ac}(x) J_i^c(z)\labe{edef} \nonumber  \\
\fraz E_i^a(x)E_i^a(x) &=& \frac{g^2}{2}\int_{x_i}^Ldz_iJ_i^b(
x_i^\perp,z_i)\int_{x_i}^Ldz_i^\prime J_i^b(x_i^\perp,z_i^\prime)\ .\labe{hej}
 \eqne
Collecting all the results the hamiltonian density reads
\eqna {\cal H}&=&\sum_i\left[\bar\psi(x)U_i(x)\right] \left[ \gamma_i
i\partial_i +m \right] \left[ U_i^{\dag}(x)\psi(x)\right]
\nonumber \\ & +& \frac{g^2}{2}
\sum_i\int_{x_i}^Ldz_i J_i^b(x_i^\perp,z_i)\int_{x_i}^Ldz_i^\prime
J_i^b(x_i^\perp,z_i^\prime) \labe{hneu}\\
&+& \frac{1}{2g^2}\sum_{ij} \tr{\left[\partial_i\left[\left(
U_i^{\dag}U_j\right)\partial_j\left(U_j^{\dag}U_i\right)\right]\right]\left[
\partial_i\left[\left(U_i^{\dag}U_j\right)
\partial_j\left(U_j^{\dag}U_i\right)\right]\right]^{\dag}}\nonumber \eqne
which is the "angle" representation we have been looking for. Note that
the locality of the Hamiltonian has been lost although we have not been
fixing the gauge yet. The electric part of the Hamiltonian is non--local
and it shows already the linear "potential" $|z_i-z_i^\prime|$ characteristic
for both the axial gauge formulation and the strong coupling limit in
lattice gauge theory. We observe also that
the Hamiltonian in QED, corresponding to eq.(\ref{hneu}), is obtained by
dropping the summations over color indices which shows the similarity
of abelian and non--abelian gauge theories in the "angle" formulation.

Finally we want to consider the form of Gauss' law operator in these
variables. We find
\eqna gf^{abc}A_i^b(x)E_i^c(x) &=& -g\left[\partial_iN_i^{ad}(x)\right]
\int dz_i\theta(z_i-x_i)\theta(x_i) J_i^d(x_i^\perp,z_i)\\
\partial_iE_i^a(x) &=& g\left[\partial_iN_i^{ad}(x)\right]
\int dz_i\theta(z_i-x_i)\theta(x_i)J_i^d(x_i^\perp,z_i)\nonumber \\
 && +gN_i^{ad}(x)\int dz_iJ_i^d(x_i^\perp,z_i)\partial_i\left[\theta(z_i-x_i)
\theta(x_i)\right]\eqne
and taking the sum of these two contributions and the charge density operator
we obtain for Gauss law the following expression
\eqa G^a(x) = -g\sum_i\left[N_i^{ad}(x)J_i^d(x)-\delta(x_i)\int dz_i
J_i^a(x_i^\perp,z_i)\right] + g\psi^{\dag}(x)\frac{\lambda^a}{2}
\psi(x)\nonumber\ . \eqe

\sect{The connection with the lattice QCD hamiltonian}
The task in this section is to show that the similarity between the variables
in our formulation and those of the lattice hamiltonian approach to QCD
can be made explicit. To this extend we replace the continuous spatial
variables
$x_i$ by lattice points $x_i^m$ and differential operators are replaced
by finite differences. In order not to violate gauge invariance, we have to
guarantee that the derivative operators only act on gauge invariant functions.
This is the case for the fermionic part of the hamiltonian.
Using the simplest prescription for the derivative operation we find
\eqa \partial_i\left[V_i^{\dag}(x)\psi(x)\right]\rightarrow\frac{1}{a^{5/2}}
\left[ V_i^{\dag}(x_i^\perp,x_i^m)\psi(x_i^\perp,x_i^m)-
V_i^{\dag}(x_i^\perp,x_i^m-a)\psi(x_i^\perp,x_i^m-a)\right]\ .\eqe
Introducing the link variables in terms of "angles" $\xi_i^m(x_i^\perp)$
\eqna L_i(x_i^{m-1},x_i^\perp) &=& P\exp\left[ig\int_{x_i^m-a}^{x_i^m}dz_i
A_i(x_i^\perp,z_i)\right]=\exp\left[i\xi_i^m(x_i^\perp)\right]
\labe{lang}\\ V_i(x_i^\perp,x_i^m) &=& L_i(x_i^{m-1},x_i^\perp)
L_i(x_i^{m-2},x_i^\perp) \dots  L_i(0,x_i^\perp)\nonumber \eqne
the fermionic hamiltonian can be cast into the familiar form
\eqna {\cal H}_{ferm}&=&\frac{i}{2a^4}\sum_i \left[
\bar\psi(x_i^\perp,x_i^{m-1})
\gamma_i L_i^{\dag}(x_i^{m-1},x_i^\perp)\psi(x_i^\perp,x_i^m)\right.
\nonumber \\ && \left. -\bar\psi(x_i^\perp,x_i^m) \gamma_i
L_i(x_i^{m-1},x_i^\perp)\psi(x_i^\perp,x_i^{m-1})\right]+
\frac{m}{a^3}\bar\psi(x_i^\perp,x_i^m)\psi(x_i^\perp,x_i^m) \labe{hferm}\
.\eqne
Defining angular momentum operators ${\cal J}_i^a(x_i^\perp,x_i^m)$
with respect to the link "angles" $\xi_i^m(x_i^\perp)$ obeying the
following commutation relations
\eqna \left[{\cal J}_i^a(x_i^\perp,x_i^m),L_j(z_j^{n-1},z_j^\perp)\right] &=&
\delta_{i,j}\delta_{n,m} \delta_{x_i^\perp,z_i^\perp}L_j(z_j^{n-1},z_j^\perp)
\frac{\lambda^a}{2}\nonumber \\ \left[{\cal J}_i^a(x_i^\perp,x_i^m),
{\cal J}_j^b(z_j^\perp,z_j^n)\right] &=& \delta_{i,j}\delta_{x_i^\perp,
z_i^\perp}\delta_{n,m}if^{abc}{\cal J}_i^c(x_i^\perp,x_i^m)
\labe{jcdef} \eqne
we arrive at the following expression for the
integral over "angular momentum" operators appearing in eq.(\ref{hneu})
\eqa \int_{x_i}^Ldz_iJ_i^c(x_i^\perp,z_i) \rightarrow \frac{1}{a^2}
\sum_{k=m}^NJ_i^c(x_i^\perp,x_i^k)= \frac{1}{a^2}N_i^{ac}(x_i^\perp,x_i^{m-1})
{\cal J}_i^c(x_i^\perp,x_i^m) \ . \eqe
The electric part of the hamiltonian eq.(\ref{hneu}) then becomes
\eqa {\cal H}_{elec}=\frac{g^2}{2a^4} \sum_i\sum_{k=m}^N\sum_{l=m}^N
J_i^a(x_i^\perp,x_i^k) J_i^a(x_i^\perp,x_i^l) = \frac{g^2}{2a^4} \sum_i
{\cal J}_i^a(x_i^\perp,x_i^m){\cal J}_i^a(x_i^\perp,x_i^m)\labe{helec} \eqe
which is the usual lattice form of this contribution
to the hamiltonian. In particular the non--locality has disappeared and the
existence of the linear "potentials" $|z_i-z_i^\prime|$ has become less
obvious. The operator which remains to be discretized is the magnetic part of
the hamiltonian. Introducing the following modification
\dia V_i^{\dag}(x_i,x_j)V_j(x_i,x_j) \rightarrow  U_{ij}^{\dag}(x_i,x_j)
= V_j^{\dag}(x_i=0,x_j)
V_i^{\dag}(x_i,x_j) V_j(x_i,x_j)V_i(x_i,x_j=0) \die
it can be rewritten such that the derivatives are acting on gauge invariant
functions. Therefore we obtain
\eqna \partial_i\left[U_{ji}(x_i,x_j)\partial_jU_{ji}^{\dag}(x_i,x_j) \right]
&\rightarrow & \frac{1}{a^2}\left[ U_{ji}(x_i-a,x_j)U_{ji}^{\dag}(x_i-a,x_j-a)
\right.\nonumber \\ && - \left. U_{ji}(x_i,x_j)U_{ji}^{\dag}(x_i,x_j-a)
\right]\labe{dij}\ . \eqne
Multiplying this result with its hermitian conjugate and using the link
matrices $L_i$ defined in eq.(\ref{lang}) the result
for the magnetic part of the Hamiltonian  eq.(\ref{hneu}) becomes
\eqna {\cal H}_{magn} &=& \frac{-1}{2g^2a^4} \sum_{i,j} \mbox{Tr}\left\{
L_i^{\dag}(x_i^{m-1},x_j^n) L_j(x_i^m,x_j^{n-1}) L_i(x_i^{m-1},x_j^{n-1})
L_j^{\dag}(x_i^{m-1},x_j^{n-1}) \right. \nonumber \\ && \left. +
L_i^{\dag}(x_i^{m-1},x_j^{n-1}) L_j^{\dag}(x_i^m,x_j^{n-1})
L_i(x_i^{m-1},x_i^n)
L_j(x_i^{m-1},x_j^{n-1}) -2\right\}
\nonumber \\ &=& \frac{1}{2g^2a^4} \sum_{ij} \tr{2-U_\Box(x_i^{m-1},x_j^{n-1})-
U_\Box^{\dag}(x_i^{m-1},x_j^{n-1})} \labe{hmag1}\eqne
where $U_\Box(x_i^{m-1},x_j^{n-1})$ is seen to be the unitary matrix obtained
from the path ordered integral along the boundary of a plaquette in the
(i,j)--plane starting from the point $(x_i^{m-1},x_j^{n-1})$. Thus collecting
the results eq.(\ref{hferm},\ref{helec},\ref{hmag1}) we find the standard
Kogut--Susskind lattice hamiltonian density \cite{Kogut}
\eqna {\cal H}_{lattice} &=& \frac{1}{a^4} \left\{\frac{g^2}{2} \sum_i
{\cal J}_i^a(x) {\cal J}_i^a(x) +\frac{1}{2g^2} \sum_{i,j}
\tr{2-U_\Box(x_i^{m-1},x_j^{n-1})- U_\Box^{\dag}(x_i^{m-1},x_j^{n-1})} \right.
\nonumber\\  && +\frac{i}{2} \sum_i \left[ \bar\psi(x_i^{\perp},x_i^{m-1})
\gamma_i L_i^{\dag}(x_i^{m-1},x_i^\perp) \psi(x_i^\perp,x_i^m)  \right.
\labe{hlatt} \\ && -\left.\left. \bar\psi(x_i^\perp,x_i^m)
\gamma_iL_i(x_i^{m-1},x_i^\perp) \psi(x_i^\perp,x_i^{m-1})
\right] +(ma) \bar\psi(x_i^\perp,x_i^m)\psi(x_i^\perp,x_i^m) \right\}
\nonumber \ .\eqne
To complete the comparison with the lattice formulation we determine also
the functional integration measure in our formulation. For this calculation
we have to evaluate the Jacobian of the transformation
from the gauge fields $A_i$ to the "angles" $\xi_j$ for which we find
\eqa {\cal R}_{ij}^{ab}(x,y) = \frac{\delta A_i^a(x)}{\delta\xi_j^b(y)}
= \frac{1}{g}\delta_{i,j} \delta^{d-1}(x_i^\perp-y_i^\perp) M_i^{bc}(y)
N_i^{ac}(x) \partial_i\delta(x_i-y_i)\ . \eqe
The Jacobian is obtained from the determinant of ${\cal R}$ which can be
written as
\eqa det({\cal R}) = det({\cal R}^{-1}) = \prod_i
det\left[M_i(z)\right] \labe{detr}\ .\eqe
The expression eq.(\ref{detr}) is just the
volume of the gauge group and after discretization this result gives the usual
integration measure in the lattice formulation (see e.g. \cite{Creutz}).

We note that in establishing the relation between the lattice and the angle
formulation of QCD an expansion in powers of the lattice spacing $a$ only
occurs due to the necessity to replace differential operators by finite
differences. As a consequence we naturally obtain the change from the flat
integration measure in the space of gauge fields $A_i(x)$ to the group
integration measure in the space of unitary matrices $V_i(x)$ or
links $L_i(x_i^m,x_i^\perp)$. It also implies that continuum operators can
be translated into lattice operators. In general this will not lead to
simple expressions in terms of plaquettes (e.g. in the case of
$\bar\psi\sigma_{\mu\nu}F^{\mu\nu}\psi$ or
$\tr{\epsilon_{\mu\nu\rho\tau}F^{\mu\nu}F^{\rho\tau}}$) unlike in the special
combination ($\tr{F_{ij}(x)F_{ij}(x)}$) appearing in the hamiltonian. With
regard to the widely discussed differences between compact and non--compact
formulations of lattice gauge theories we clearly observe that the existence
of the strong coupling limit is related to the replacement of an unbounded
by a bounded magnetic operator
\dia U_{ji}(x_i,x_j)\partial_j U_{ji}^{\dag}(x_i,x_j)\rightarrow
\left|\left| \frac{1}{a}\left[1- U_{ji}(x_i,x_j) U_{ji}^{\dag}(x_i,x_j-a)
\right] \right|\right|  \leq \frac{2}{a}\ . \die
This suggests to use a non--compact continuum formulation, separate gauge
variant degrees of freedom and introduce a lattice for the gauge invariant
degrees of freedom as a procedure leading to a meaningful non--compact lattice
formulation of non--abelian gauge theories.

\sect{Summary and conclusion}
In summary we have arrived at a continuum formulation of non--abelian gauge
theories
entirely in terms of angular degrees of freedom. The objective for doing so was
the wish to have a formulation that leaves us freedom in choosing appropriate
unphysical "angle" variables. This may prove importantfor being able to
develop useful
approximations to understand the low energy properties of QCD. The resulting
formulation can be shown to be equivalent in a finite volume to Bars corner
variable formulation and is like Bars formulation \cite{Bars1} closely
related to the lattice hamiltonian approach to QCD \cite{Kogut} as we
have shown explicitly. In addition to this it is a formulation which has
the built--in freedom in selecting unphysical variables. Furthermore it has
been shown for the axial gauge representation of QCD \cite{Lenz}
already that this reformulation leads to great technical simplifications in
the elimination of unphysical variables \cite{Stoll}. All these
advantages together may render the "angle" formulation an useful starting
point for investigating non--perturbative aspects of QCD in terms of gauge
invariant degrees of freedom.
\vspace{.1cm}\\
\begin{center} {\large \bf Acknowledgements}\end{center}
We thank Profs. K.Ohta, M.Thies and K.Yazaki for helpful
discussions and acknowledge gratefully the financial support by the
"Japan Society for the Promotion of Science".

\end{document}